\documentclass[fleqn,10pt]{wlscirep}
\usepackage[utf8]{inputenc}
\usepackage[T1]{fontenc}
\usepackage{graphicx}
\usepackage{subcaption}
\usepackage{geometry}
\usepackage{amssymb} 
\usepackage{pifont} 
\usepackage{multirow}
\usepackage{enumitem}
\usepackage{svg}
\usepackage{booktabs}
\usepackage{url}
\usepackage[numbers,sort&compress,super]{natbib}

\title{Open Synthetic University Assignment Rubric and Submission Dataset for LLM Educational Feedback Research}

\author{Keyang Qian}
\author{Kaixun Yang}
\author{Wei Dai}
\author{Flora Jin}
\author{Yixin Cheng}
\author{Rui Guan}
\author{Sadia Nawaz}
\author{Zachari Swiecki}
\author{Guanliang Chen}
\author[2,*]{Lixiang Yan}
\author[3,1,*]{Dragan Ga\v{s}evi\'{c}}

\affil{Centre for Learning Analytics, Faculty of Information Technology, Monash University, Melbourne, VIC, Australia}
\affil[2]{School of Education, Tsinghua University, Beijing, PR China}
\affil[3]{Faculty of Education and School of Computing \& Data Science, The University of Hong Kong, Hong Kong SAR, PR China}

\affil[*]{lixiangyan@mail.tsinghua.edu.cn, dgasevic@hku.hk}

\begin{abstract}
Using Large Language Models (LLMs) to provide educational feedback to students on their assignments has attracted significant attention in the Artificial Intelligence in Education (AIED) field. Yet, there is currently no large-scale open-source dataset of student assignments that includes detailed assignment descriptions, rubrics, and student submissions across diverse courses. As a result, empirical research on automated methodologies for generating effective educational feedback remains constrained by data availability. In this paper, we introduce a large-scale synthetic computer science university assignment dataset for LLM-based educational feedback research, called SCALEFeedback (Synthetic Computer science Assignments for LLM Educational Feedback Research). The dataset is generated via the Sophisticated Assignment Mimicry (SAM) framework, specifically designed to synthesize educational artifacts through one-to-one LLM-based imitation of authentic assignment descriptions, rubrics, and student submissions, aiming to minimize student privacy and intellectual property risks associated with releasing raw educational artifacts. Our open-source dataset contains 10,000 synthetic student submissions spanning 155 assignments across 59 university-level computer science courses. Technical validation demonstrates that the synthetic dataset closely approximates the linguistic characteristics and macro-score distributions of the source data while effectively preventing the leakage of direct personal identifiers under automated verification. The release of this dataset is a valuable contribution to researchers who aim to develop LLM-based methods for offering high-quality, automated computing education assignment feedback in a scalable way.
\end{abstract}
\begin{document}

\maketitle

\section*{Background \& Summary}

Feedback plays a crucial role in student learning and academic success \citep{m:3}, but providing timely, high-quality, and personalized feedback can be challenging, particularly in large computing classes where student enrollment is increasingly outpacing instructional resources \citep{n:1,n:2}. The integration of Large Language Models (LLMs) into educational settings, particularly for generating automated formative and summative feedback on student assignments, has emerged as a major area of research \citep{k:1,r:1,d:1,qian2025towards,qian2025dean}, especially in computer science education \citep{d:2,b:1,h:1,l:1,l:2}.

Despite this promise, prior LLM-based feedback research has largely focused on isolated experiments involving specific programming tasks \citep{p:1} or single university courses \citep{l:1}, limiting the broader assessment of these techniques across diverse curricular contexts. A critical barrier limiting scalable research is the scarcity of open, shareable educational datasets that encompass diverse student work \citep{g:1,s:1,m:1,m:2,w:1}. To the best of our knowledge, there is currently no publicly available dataset that comprehensively includes authentic assignment descriptions, grading rubrics, and diverse student submissions across a wide range of computing topics. The direct public release of such materials is fundamentally constrained by institutional intellectual property rights and student data privacy regulations.

One promising approach for addressing this data scarcity is large-scale synthetic data generation using LLMs. Prompting LLMs to synthesize benchmark data has gained traction across numerous domains, including mathematics \citep{t:1,y:1}, science \citep{l:3}, medicine \citep{h:2,h:3,s:2}, law \citep{y:2,z:1}, and education \citep{d:3,l:4,l:5}. However, existing synthetic educational datasets rarely include aligned assignment descriptions, grading rubrics, and complete student assignment submissions.

Synthesis approaches that capture key distributional characteristics of real educational data while addressing privacy and copyright risks are therefore needed. In response, we constructed a large-scale open-source dataset of student assignments -- named the SCALEFeedback dataset (Synthetic Computer science Assignments for LLM Educational Feedback Research) -- that includes detailed synthetic assignment descriptions, rubrics, and student submissions across 59 computer science courses. The dataset was generated using the \textbf{S}ophisticated \textbf{A}ssignment \textbf{M}imicry (\textit{SAM}) framework, designed with the goal of mitigating privacy risks and copyright concerns through one-to-one synthetic imitation. By abstracting sensitive source artifacts into fictionalized contexts, this approach approximates the underlying semantic structures and submission characteristics while aiming to eliminate personally identifiable information. Consequently, the dataset can be shared openly under a permissive license to support discovery, methodological benchmarking, and tool development, while mitigating the primary legal and ethical barriers associated with releasing authentic student work.

Our main contributions are summarized as follows:
\begin{enumerate}[label=\textbf{\textperiodcentered}]
  \item We release an open assignment dataset, SCALEFeedback, containing 10,000 synthetic student submissions spanning 155 assignments across 59 university-level computer science courses, designed to support benchmarking and development of automated LLM-based feedback methods in computing education.
  \item We report extensive technical validation demonstrating that the synthetic dataset closely resembles the linguistic metrics, readability metrics, and macro-level mark distributions of the source data, while verifying the removal of explicit student identifiers under an automated evaluation protocol.
  \item We show that the dataset serves as an initial automated baseline proxy for studies of LLM-based educational feedback generation, demonstrating high alignment across automated feedback evaluation attributes when comparing feedback generated for synthetic submissions against feedback generated for corresponding real submissions.
  \item We document the SAM one-to-one LLM-based synthesis framework, providing a transparent and reproducible workflow for institutional educational data synthesis and privacy-oriented risk mitigation.
\end{enumerate}

\section*{Methods}

\subsection*{Real-world Dataset Acquisition}
Following approval from the Monash University Human Research Ethics Committee (Project ID: 29874), we retrieved the assignment archives of all undergraduate and postgraduate coursework units delivered during the first semester of 2021 by the Faculty of Information Technology at Monash University. We confirm that all data collection and processing procedures were performed in accordance with relevant institutional guidelines and ethical regulations. From these courses, we collected complete assessment packages, including assignment briefs, problem descriptions, grading rubrics, background reading materials, and authentic student submissions. 

The source assignment materials were originally submitted in diverse file formats, including \texttt{.pdf}, \texttt{.txt}, \texttt{.sql}, \texttt{.py}, \texttt{.docx}, \texttt{.ipynb}, and \texttt{.rmd}\footnote{The file extensions correspond to: \texttt{.rmd} (R Markdown dynamic documents combining R code and narrative text), \texttt{.ipynb} (Jupyter Notebooks containing executable code blocks and markdown), \texttt{.py} (Python source code scripts), \texttt{.sql} (Structured Query Language database scripts), and \texttt{.docx}, \texttt{.pdf}, \texttt{.txt} (standard word processing and plain text document formats).}. To prepare these materials for processing across standard LLM APIs, all non-plain-text files were converted into UTF-8 encoded text files. Non-textual structural elements, such as tables and formatted rubric grids, were serialized into structured textual words and markdown representations. 

\noindent\textbf{Structural Reduction and Composite Assignment Modalities.} We explicitly acknowledge that converting multi-modal and structured programming artifacts (such as \texttt{.py}, \texttt{.sql}, and \texttt{.ipynb}) into flat text files involves a degree of structural reduction. Specifically, flattening text may obscure hierarchical logic, spatial indentation, database schema visual relationships, and notebook cell execution boundaries. In authentic computer science education, assignments and student submissions are predominantly composite and hybrid in nature: a single submission frequently combines natural language problem analysis, architectural diagrams or tabular schemas, embedded executable scripts, and reflective discussions. Because these multi-part artifacts interweave narrative and code in varying proportions, imposing a rigid, post-hoc categorization into isolated task types (e.g., purely narrative versus purely code) is methodologically arbitrary. We therefore preserve the holistic corpus distribution during technical validation while explicitly noting structural flattening as a known limitation.

\noindent\textbf{Context-Length Filtering and Sampling.} Across the 59 courses, a total of 45,126 authentic student submissions were collected. To ensure that the resulting dataset remains accessible for benchmarking across a wide range of LLMs without exceeding context window constraints, we applied an upper-bound length cutoff. Specifically, we excluded assignments where the combined length of the assignment description, rubric, and student submission exceeded 7,500 words ($\approx 10,000$ LLM tokens). This cutoff ensured compatibility with standard context windows (e.g., 16k tokens in models such as GPT-3.5-Turbo \citep{OpenAI2025GPT35Turbo}). We note that this filtering step introduces a deliberate sampling boundary: it primarily retains foundational to intermediate computer science coursework tasks, while excluding long, multi-file software engineering deliverables and capstone projects. 

From the filtered pool, we randomly sampled 10,000 student submissions and their corresponding assignment specifications, proportionally weighted by the number of submissions per assignment. These 10,000 submissions span 155 unique assignments across 59 computing units.

\begin{figure}[ht]
  \centering
    \includegraphics[width=\linewidth]{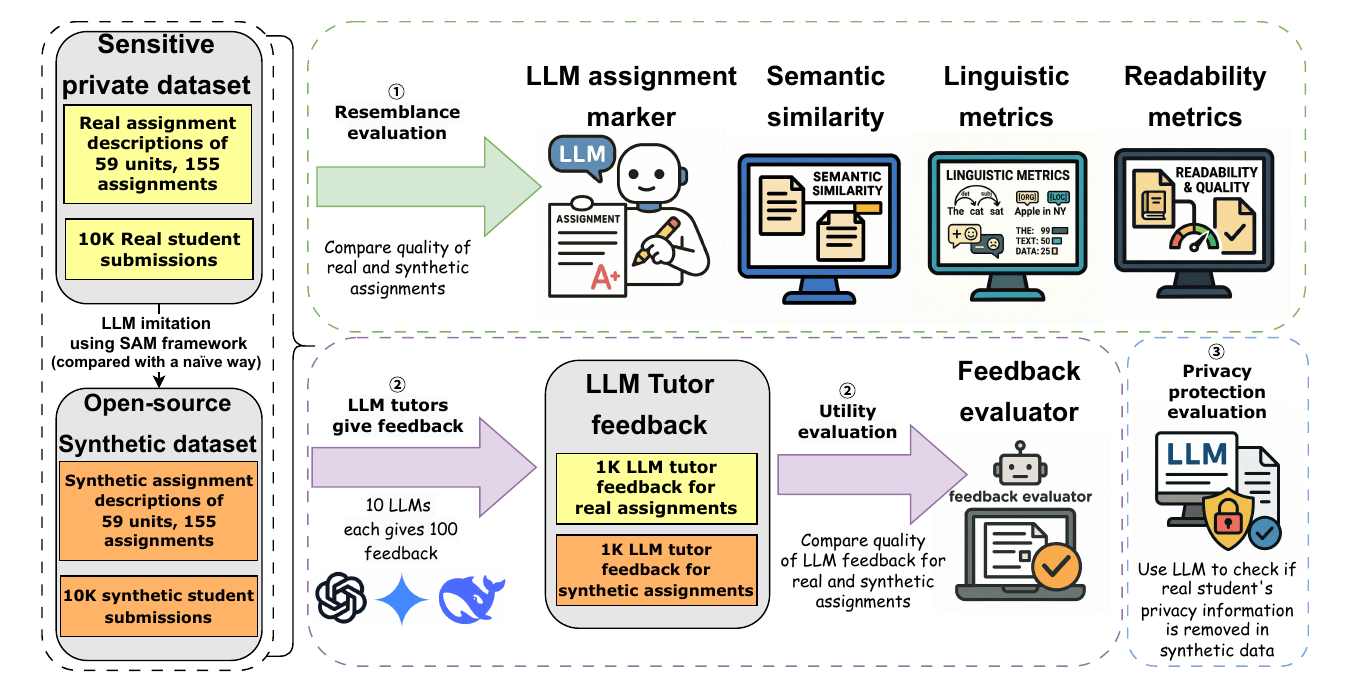}
      \caption{Overview of dataset synthesis and technical validation workflow.}
      \label{fig:overview}
\end{figure}

\subsection*{Data Synthesis via Sophisticated Assignment Mimicry (SAM)}
To generate synthetic assignment specifications, rubrics, and student submissions, we developed the \textbf{S}ophisticated \textbf{A}ssignment \textbf{M}imicry (\textit{SAM}) framework (Figure~\ref{fig:overview}). The SAM framework employs structured prompt-chaining and self-reflection loops to emulate the pedagogical intent, problem complexity, rubric criteria, and student error distributions of the source data within entirely fictionalized contexts.

\noindent\textbf{1. Synthetic Assignment Description and Rubric Generation.} The generator model adopts the persona of a university professor. Given an authentic assignment description, the model executes a three-step generation and reflection protocol:
\begin{enumerate}[label=\arabic*)]
  \item \textbf{Decomposition:} The model extracts the core pedagogical themes, learning objectives, problem difficulty, narrative style, structural sections, and detailed rubric grading tiers of the authentic assignment.
  \item \textbf{Contextual Re-synthesis:} The model generates a novel assignment description by transposing the extracted dimensions into an entirely new scenario, domain context, variable naming scheme, and dataset structure, while calibrating the output to match the target word count of the original assignment.
  \item \textbf{Self-Verification and Regeneration Loop:} The model inspects its generated assignment against the original requirements. If the pedagogical objectives or task complexity diverge, or if the generated word count deviates significantly from the target length, the generation is discarded and regenerated.
\end{enumerate}

\noindent\textbf{2. Synthetic Student Submission Generation.} To generate corresponding synthetic submissions, the generator model adopts the persona of a virtual student. The model receives a paired input: the newly generated synthetic assignment (Assignment A), the authentic assignment brief (Assignment B), and an authentic student submission for Assignment B. The generation operates via a four-step calibration pipeline:
\begin{enumerate}[label=\arabic*)]
  \item \textbf{Cross-Assignment Goal Alignment:} The model maps the sub-tasks, problem questions, and rubric criteria of Synthetic Assignment A to those of Authentic Assignment B.
  \item \textbf{Authentic Submission Evaluation:} The model evaluates the authentic student submission against the Assignment B rubric, diagnosing the exact level of completeness, depth of explanation, specific conceptual errors, and the resulting mark tier.
  \item \textbf{Fictionalized Imitation:} The model synthesizes a virtual student submission answering Assignment A. It is instructed to emulate the identical quality tier, specific patterns of misconceptions, and technical depth evaluated in Step 2, calibrated to the word length of the original submission.
  \item \textbf{Rubric Self-Grading and Score Calibration:} The model evaluates its own synthesized submission against the Assignment A rubric. If the estimated mark of the synthetic submission deviates from the authentic mark awarded in Step 2, the model iteratively revises the synthetic submission until the rubric marks closely converge.
\end{enumerate}

\noindent\textbf{Fictional Metadata and In-Prompt Privacy Handling.} In response to privacy requirements, the SAM synthesis prompts strictly instruct the generator model to exclude all authentic personal identifiers (e.g., student names, identification numbers, staff names, institutional email addresses, and institutional URLs). If an assignment format requires student identification headers, the prompt instructs the model to either omit the header entirely or substitute randomized, fictional placeholder names and dummy alphanumeric strings (e.g., ``Student\_ID: 98765432'').

\noindent\textbf{External LLM Privacy Gate.} To reinforce privacy preservation, we instituted an external, multi-stage LLM-based privacy filter. An independent reasoning model (o4-mini \citep{OpenAI2025o4mini}, high reasoning effort) cross-checks the extracted real personal identifiers against the generated candidate. If any genuine identifier from the source text is detected in the synthetic text, the candidate submission is permanently discarded and regenerated.

\noindent\textbf{LLM Model Specifications.} For generating synthetic assignment descriptions and rubrics, we utilized OpenAI o3-pro (high reasoning effort) \citep{OpenAI2025o3pro}. For generating synthetic student submissions, we utilized OpenAI o4-mini (high reasoning effort) \citep{OpenAI2025o4mini} to balance synthesis fidelity and computational cost. All models were accessed via enterprise API endpoints with explicit data-logging exemptions, ensuring that neither source nor synthetic data were stored or utilized for future model training iterations.

\subsection*{Illustrative Transformation Case Study}
To concretely demonstrate the one-to-one transformation executed by the SAM framework, we illustrate a representative computer science assignment from our corpus:
\begin{itemize}[label=\textbf{\textperiodcentered}]
  \item \textbf{Assignment-Level Transformation:} An authentic assignment (Assignment B) focusing on relational database design and SQL query optimization for an e-commerce warehouse (involving normalization to 3NF, multi-table joins, and indexing) is processed. The SAM framework re-contextualizes this into a synthetic municipal transit scheduling system (Assignment A). The synthetic specification establishes an entirely new schema (routes, bus maintenance logs, passenger taps) while preserving the identical pedagogical objectives, cognitive difficulty, and rubric weighting (schema normalization 30\%, query formulation 40\%, indexing rationale 30\%).
  \item \textbf{Submission-Level Calibration:} An authentic student submission for Assignment B demonstrated partial mastery (awarded 65/100): the student successfully wrote basic relational queries but incorrectly applied an \texttt{INNER JOIN} instead of a \texttt{LEFT OUTER JOIN} for optional customer orders, and omitted the indexing explanation. Following the four-step submission pipeline, SAM evaluates these specific errors, maps them to Assignment A, and synthesizes a transit-themed submission that exhibits an analogous conceptual flaw (e.g., an incorrect join condition omitting routes without maintenance logs and missing index justifications). The synthetic self-grading step ensures the generated submission achieves a matching 65\% mark tier without reproducing any verbatim text from the student.
\end{itemize}

\subsection*{Baseline Comparison Generation}
To evaluate the efficacy of the SAM framework, we implemented a \textit{Naïve Mimicry} baseline. This baseline replaced the multi-step structured prompting with a single zero-shot instruction: ``\textit{Please generate a synthetic university assignment description (or student submission) by imitating the following assignment description (or submission)}'', without intermediate rubric evaluation steps or the external privacy gate. The identical base models were used for fair comparison.

\section*{Data Records}

The full synthetic dataset is publicly accessible via the Open Science Framework (OSF) repository \citep{dataset} at \url{https://doi.org/10.17605/OSF.IO/9PWMX}. The dataset is released under a Creative Commons Attribution 4.0 International license (CC BY 4.0) and provided as a single tabular CSV file (\texttt{synthetic\_assignments.csv}) comprising 10,000 rows. The dataset includes the following fields:

\begin{enumerate}[label=\textbf{\textperiodcentered}]
    \item \textbf{unit\_name}: The anonymized designation of the academic course unit (spanning 59 computer science units).
    \item \textbf{assignment\_name}: An identifier code for the assessment task (155 unique assignments across all units).
    \item \textbf{Virtual assignment description}: Text containing the synthetic assignment description, learning objectives, problem context, and grading rubrics.
    \item \textbf{Virtual student submission}: Text containing the synthetic student submission. For multi-part assignments, this contains the synthetic counterpart of the primary submitted text artifact.
    \item \textbf{Assignment Type}: Categorical classification tags indicating task nature (e.g., `Coding Task', `Report Writing', `Data Analysis', `System Design', `Project Management', `Test/Exam'). Descriptive statistics are shown in Table~\ref{tab:assignment_types}.
\end{enumerate}

\begin{table}[ht]
\centering
\begin{tabular}{lcc}
\hline
\textbf{Assignment Type} & \textbf{Assignment Descriptions Count} & \textbf{Assignment Submissions Count} \\ \hline
Report Writing           & 74   & 4154         \\ 
Coding Task              & 56   & 5708         \\ 
Data Analysis            & 26   & 1381         \\ 
System Design            & 18   & 1183         \\ 
Project Management       & 17   & 618          \\ 
Test/Exam                & 3    & 155          \\ \hline
\end{tabular}
\caption{Descriptive statistics of assignment types in the SCALEFeedback corpus.}
\label{tab:assignment_types}
\end{table}

Descriptive statistics of estimated token counts for synthetic assignments when processed by LLMs are provided in Table~\ref{tab:token_summary}. The maximum total length of 7,410.7 tokens remains well within standard LLM context windows (e.g., 16,385 tokens for GPT-3.5-Turbo \citep{OpenAI2025GPT35Turbo}).

\begin{table}[ht]
\centering
\begin{tabular}{lccc}
\toprule
\textbf{Metric} & \textbf{Assignment Description Tokens} & \textbf{Assignment Submission Tokens} & \textbf{Sum Tokens} \\ \midrule
Mean            & 1930.9                                & 1022.8         & 2953.7                      \\
Max             & 3996.0                                & 4740.0         & 7410.7                      \\ \bottomrule
\end{tabular}
\caption{Descriptive statistics of estimated LLM token lengths for synthetic assignments, calculated using the standard approximation of 100 tokens $\approx$ 75 words \citep{OpenAI2025Tokens}. `Sum Tokens' represents the combined token count of an assignment description and its corresponding student submission.}
\label{tab:token_summary}
\end{table}

\section*{Technical Validation}

From the 155 assignments, we sampled a representative subset of 500 paired synthetic student submissions and assignment descriptions, proportionally weighted by assignment volume, to conduct technical resemblance validation. This sample size yields a margin of error of $\pm 4.4\%$ at a 95\% confidence level.

\subsection*{Resemblance Evaluation}
We evaluated synthetic fidelity across five primary dimensions:
\begin{enumerate}[label=\textbf{\textperiodcentered}]
  \item \textbf{BERTScore F1 and Std:} Semantic similarity between real and synthetic pairs evaluated via BERTScore \citep{z:4}.
  \item \textbf{PCC and MAE for Length:} Pearson Correlation Coefficient (PCC) and Mean Absolute Error (MAE) of word counts.
  \item \textbf{PCC and MAE for Assignment Marks:} Grade preservation assessed by LLM graders (averaging scores from o3-high, o4-mini-high, and GPT-4.1) \citep{y:3,z:5} across matching rubrics.
  \item \textbf{Cosine Similarity of LIWC Metrics:} Linguistic Inquiry and Word Count (LIWC-22) \citep{p:5} profiles comprising 83 stylistic and psychological linguistic features.
  \item \textbf{Cosine Similarity of Coh-Metrix Indices:} Readability and syntactic complexity evaluated across 68 Coh-Metrix indices \citep{m:4}.
\end{enumerate}

\begin{figure}[ht]
  \centering
    \includegraphics[width=0.7\linewidth]{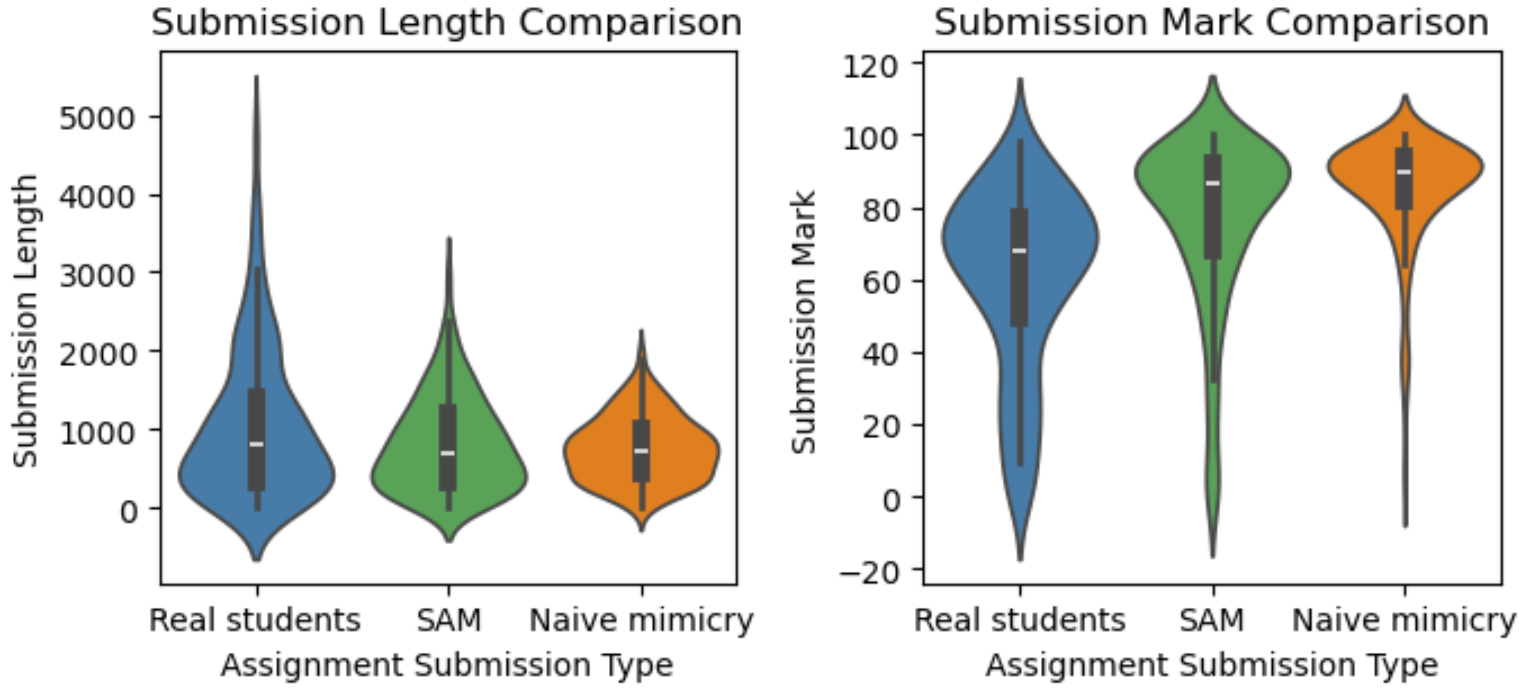}
      \caption{Distributional comparisons of submission length and marks between authentic student submissions and submissions generated by SAM and naïve mimicry.}
      \label{fig:submission_violin}
\end{figure}

\begin{figure}[ht]
    \centering
    \begin{subfigure}[b]{0.49\textwidth}
        \centering
        \includegraphics[width=\textwidth]{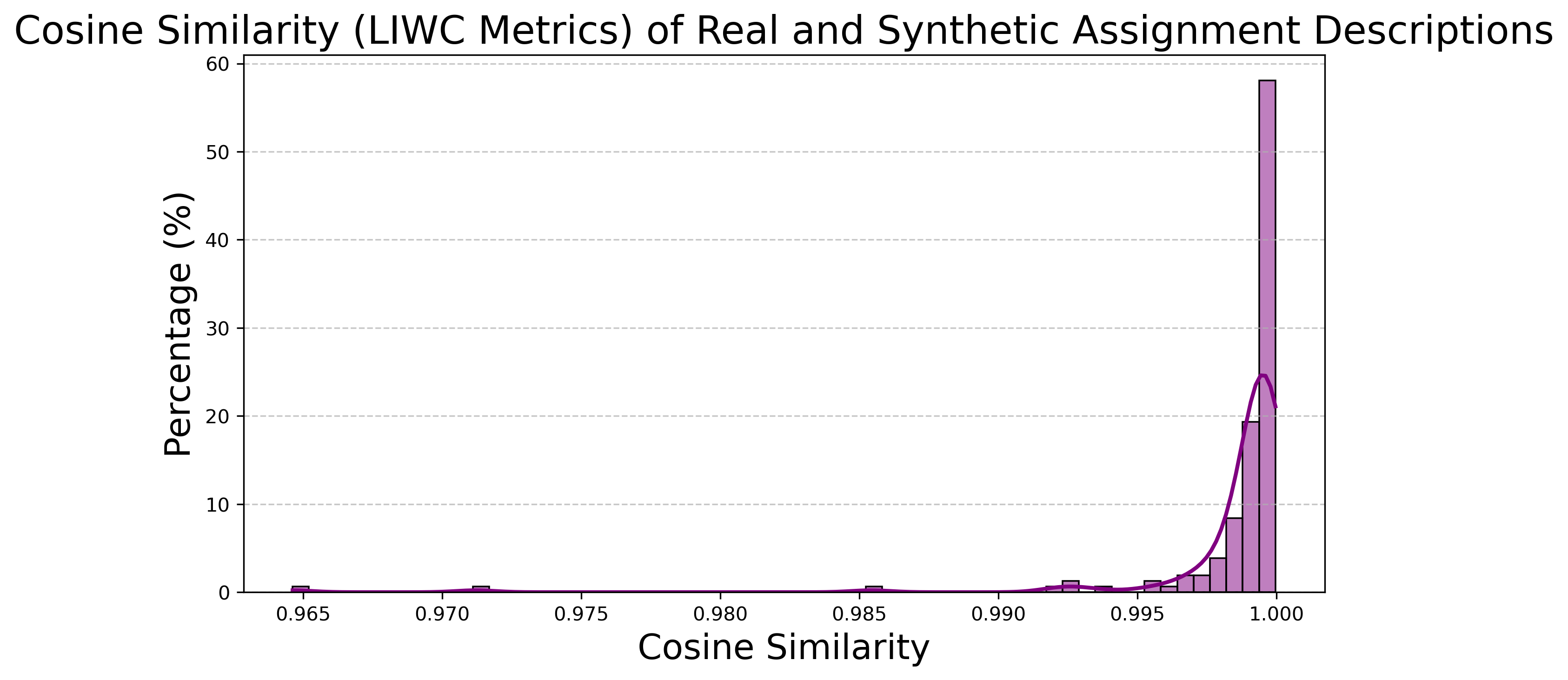} 
        \caption{}
        \label{fig:cosine_similarity_liwc_descriptions}
    \end{subfigure}
    \hfill
    \begin{subfigure}[b]{0.49\textwidth}
        \centering
        \includegraphics[width=\textwidth]{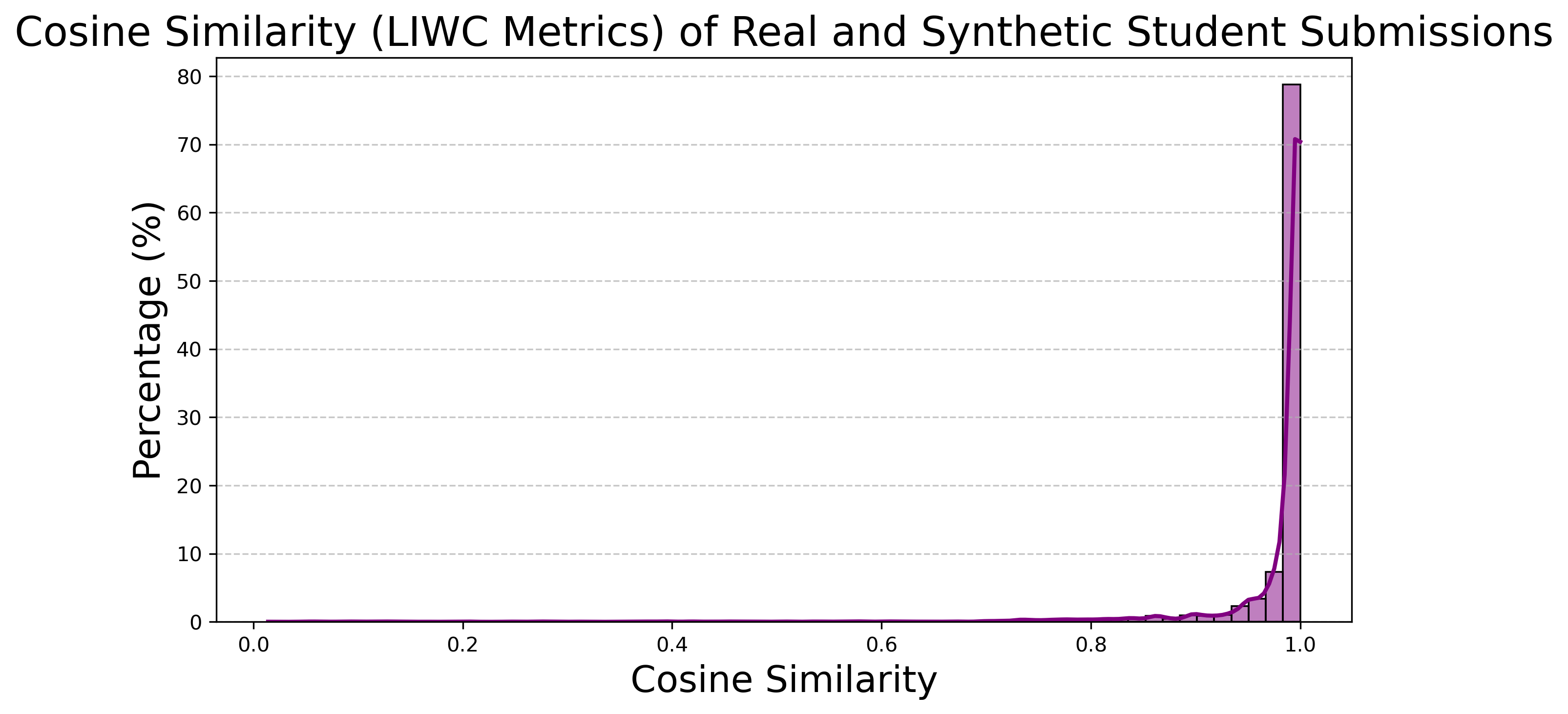}
        \caption{}
        \label{fig:cosine_similarity_liwc_submissions}
    \end{subfigure}
    \vspace{1em}
    \begin{subfigure}[b]{0.49\textwidth}
        \centering
        \includegraphics[width=\textwidth]{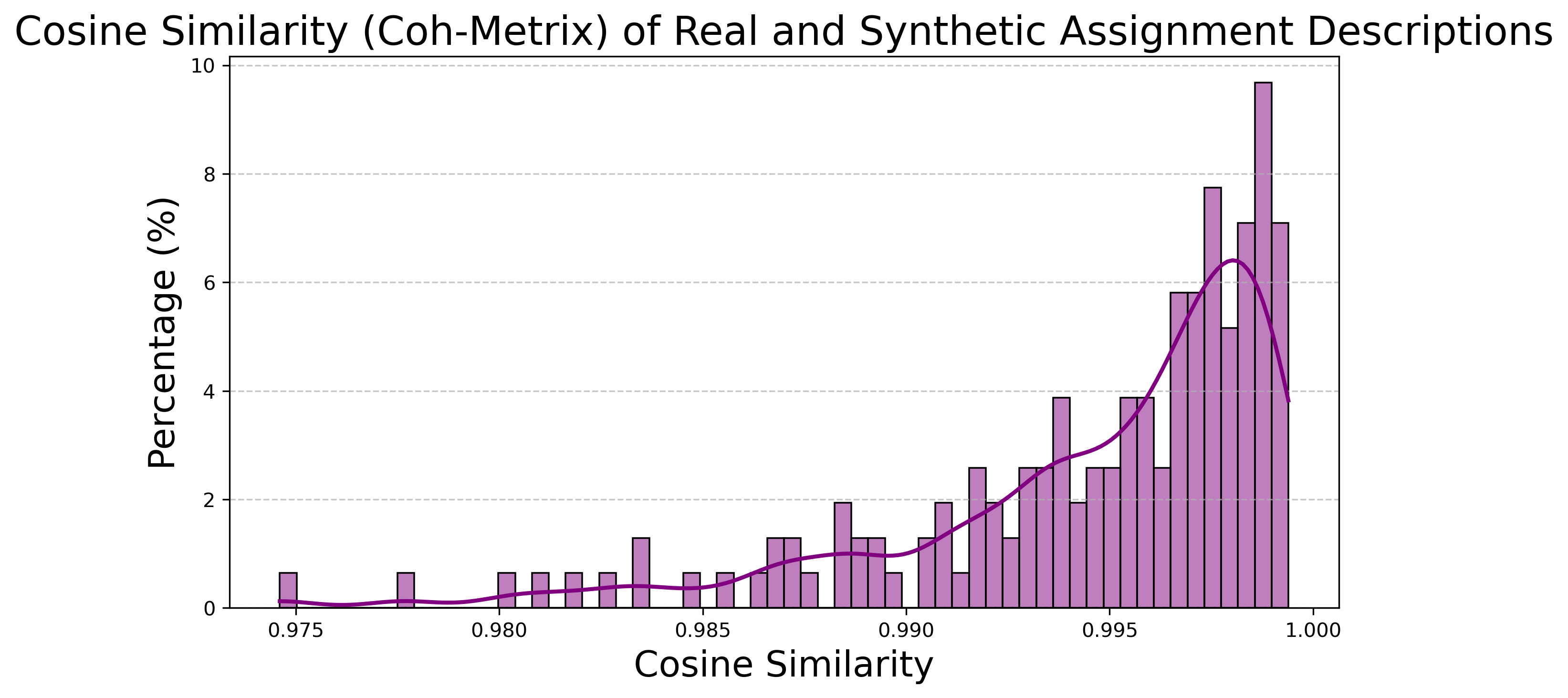}
        \caption{}
        \label{fig:cosine_similarity_coh_descriptions}
    \end{subfigure}
    \hfill
    \begin{subfigure}[b]{0.49\textwidth}
        \centering
        \includegraphics[width=\textwidth]{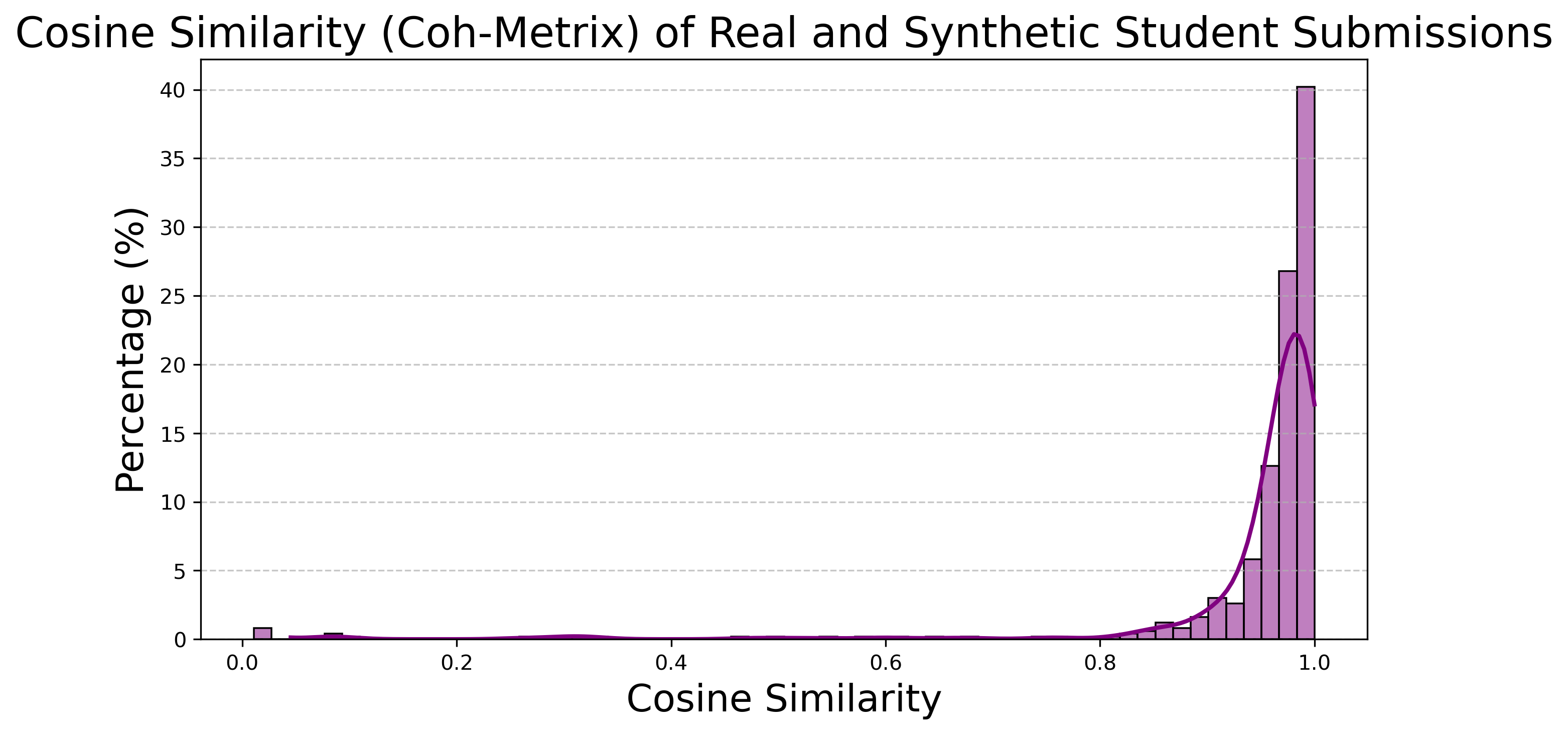}
        \caption{}
        \label{fig:cosine_similarity_coh_submissions}
    \end{subfigure}
    \caption{Distributions of cosine similarities for LIWC and Coh-Metrix metrics across descriptions and submissions.}
    \label{fig:linguistic_metrics}
\end{figure}

The resemblance results are summarized in Tables~\ref{tab:resemblance}--\ref{tab:linguistic_metrics} and Figures~\ref{fig:submission_violin}--\ref{fig:linguistic_metrics}. 

\noindent\textbf{Semantic and Length Correlation.} As shown in Table~\ref{tab:resemblance}, synthetic assignment descriptions generated via SAM achieved a length PCC of 0.931 ($\text{MAE} = 586.65$ words), substantially outperforming naïve mimicry ($\text{PCC} = 0.656$, $\text{MAE} = 1018.21$). For student submissions, SAM achieved a length PCC of 0.852 ($\text{MAE} = 335.43$) and BERTScore F1 of 0.840 ($\text{Std} = 0.020$).

\begin{table}[ht]
\centering
\resizebox{\linewidth}{!}{%
\begin{tabular}{ccccccc}
\hline
\textbf{Synthetic Data} & \textbf{BERTScore F1} & \textbf{Std} & \textbf{PCC for Length} & \textbf{MAE (Words)} & \textbf{PCC for Marks} & \textbf{MAE (Marks)} \\ \hline
\textbf{Assignment Descriptions} & & & & & & \\
Naïve Mimicry & \textbf{0.875} & 0.028 & 0.656 & 1018.21 & -- & -- \\
SAM & 0.859 & 0.038 & \textbf{0.931} & \textbf{586.65} & -- & -- \\ \hline
\textbf{Student Submissions} & & & & & & \\
Naïve Mimicry & 0.819 & 0.036 & 0.598 & 524.86 & 0.421 & 26.09 \\
SAM (Overall) & \textbf{0.840} & 0.020 & \textbf{0.852} & \textbf{335.43} & \textbf{0.624} & \textbf{19.98} \\ \hline
\end{tabular}%
}
\caption{Resemblance evaluation results of synthetic assignment descriptions and student submissions compared against authentic data across the complete corpus. PCC denotes Pearson Correlation Coefficient; MAE denotes Mean Absolute Error.}
\label{tab:resemblance}
\end{table}

\noindent\textbf{Mark Distributions and Tail Error Breakdown.} For assignment marks, submissions generated by SAM showed a moderate positive correlation with authentic marks ($\text{PCC} = 0.624$, $\text{MAE} = 19.98$), outperforming naïve mimicry ($\text{PCC} = 0.421$, $\text{MAE} = 26.09$). However, as illustrated in the violin plots (Fig.~\ref{fig:submission_violin}), the SAM distribution exhibits noticeable \textit{regression to the mean}, concentrating more heavily around median performance bands. 

A granular stratification of mark MAE across performance tiers confirms that this error is heavily concentrated in the lower tail:
\begin{itemize}[label=\textbf{\textperiodcentered}]
  \item For passing and high-performing submissions scoring $\ge 60/100$ ($N = 125$, 62.5\% of the sample), the MAE is relatively modest at 14.21 marks, dropping to 9.74 marks for high-distinction submissions ($80$--$100$) with negligible bias ($+0.30$).
  \item Conversely, for failing submissions scoring $< 40/100$ ($N = 41$, 20.5\% of the sample), the MAE increases substantially to 33.11 marks ($\text{Std} = 27.93$), accompanied by a positive bias of $+29.89$ marks (authentic mean: 18.17 vs. synthetic mean: 48.06).
\end{itemize}
This discrepancy arises because the generator LLMs tend to produce syntactically functional, coherent responses and struggle to synthesize severe conceptual breakdowns or bizarre formatting failures. Consequently, the dataset approximates macro-level score distributions for typical submissions but does not faithfully replicate extreme lower-tail student behaviors.

\noindent\textbf{Linguistic and Readability Profiles.} Table~\ref{tab:linguistic_metrics} and Figure~\ref{fig:linguistic_metrics} present the cosine similarities for LIWC and Coh-Metrix metrics. Both synthetic descriptions and submissions exhibited $>0.9$ cosine similarity with real counterparts across the 83 LIWC features and 68 Coh-Metrix indices. We emphasize that these high scores demonstrate strong preservation of \textit{stylistic, linguistic, and syntactic readability characteristics}, but should not be conflated with cognitive or pedagogical equivalence.

\begin{table}[ht]
\centering
\begin{tabular}{ccc}
\hline
\textbf{Component} & \textbf{LIWC Metrics Cosine Similarity} & \textbf{Coh-Metrix Cosine Similarity} \\ \hline
Assignment Descriptions & 0.999 (0.998, 0.999) & 0.995 (0.994, 0.995) \\
Student Submissions & 0.976 (0.975, 0.977) & 0.944 (0.932, 0.956) \\ \hline
\end{tabular}
\caption{Cosine similarities (with 95\% confidence intervals) of linguistic and readability features between authentic and synthetic artifacts.}
\label{tab:linguistic_metrics}
\end{table}

\subsection*{Utility Evaluation as an Automated Feedback Proxy}

To assess the utility of SCALEFeedback as a proxy for automated feedback research, we evaluated whether feedback generated on synthetic submissions triggers the same pedagogical components as feedback generated on matching real submissions. We employed 10 commercial LLMs across three model families: OpenAI (o3-2025-04-16, o4-mini-2025-04-16, gpt-4.1-2025-04-14, gpt-4.1-nano-2025-04-14) \citep{OpenAI2025GPT41}, Gemini (gemini-2.5-pro, gemini-2.5-flash, gemini-2.0-flash-001, gemini-2.0-flash-lite-001) \citep{GeminiTeam2024}, and DeepSeek (DeepSeek V3 0324, DeepSeek R1 0528) \citep{DeepSeek2024V3,DeepSeek2025R1}. Each model generated feedback for 100 paired real and synthetic submissions, yielding 2,000 feedback samples.

Feedback quality was evaluated using Aldino et al.'s automated feedback framework \citep{a:2}, which scores learner-centered feedback across seven pedagogical dimensions. Treating feedback attributes generated on real submissions as the reference standard, we measured alignment, sensitivity, specificity, and F1 scores (Table~\ref{tab:feedback_similarity}). Across all attributes, we observed an average alignment of 0.847 (95\% CI: [0.838, 0.855]) and an F1 score of 0.856 (95\% CI: [0.848, 0.865]). These results indicate that synthetic submissions provide high automated alignment with real data across these dimensions, supporting their role as an initial automated baseline proxy for feedback systems research.

\begin{table}[ht]
\centering
\resizebox{\linewidth}{!}{%
\begin{tabular}{lllll}
\toprule
\textbf{Attribute} & \textbf{Alignment (95\% CI)} & \textbf{Sensitivity (95\% CI)} & \textbf{Specificity (95\% CI)} & \textbf{F1 Score (95\% CI)} \\
\midrule
Upcoming Similar Tasks & 0.983 (0.974, 0.990) & 0.989 (0.983, 0.995) & 0.000 (0.000, 0.000) & 0.991 (0.987, 0.995) \\
Strengths and Weaknesses & 0.998 (0.995, 1.000) & 1.000 (1.000, 1.000) & 0.333 (0.000, 1.000) & 0.999 (0.997, 1.000) \\
Performance Summary & 0.774 (0.748, 0.799) & 0.887 (0.866, 0.908) & 0.304 (0.243, 0.376) & 0.864 (0.846, 0.880) \\
Meeting Learning Objective & 0.999 (0.997, 1.000) & 0.000 (0.000, 0.000) & 0.999 (0.997, 1.000) & 0.000 (0.000, 0.000) \\
Active Role & 0.695 (0.669, 0.724) & 0.471 (0.421, 0.527) & 0.825 (0.796, 0.855) & 0.531 (0.483, 0.577) \\
Affirmation and Encouragement & 0.707 (0.678, 0.736) & 0.571 (0.512, 0.628) & 0.756 (0.726, 0.788) & 0.509 (0.461, 0.557) \\
Student Teacher Relationship & 0.771 (0.743, 0.796) & 0.576 (0.520, 0.629) & 0.859 (0.830, 0.886) & 0.610 (0.564, 0.658) \\
\textbf{Average} & \textbf{0.847 (0.838, 0.855)} & \textbf{0.855 (0.843, 0.866)} & \textbf{0.837 (0.825, 0.849)} & \textbf{0.856 (0.848, 0.865)} \\
\bottomrule
\end{tabular}
}
\caption{Utility reliability metrics evaluating synthetic assignments as baseline proxies in LLM-based educational feedback research. Alignment denotes attribute agreement between feedback on synthetic submissions and real submissions. Sensitivity, Specificity, and F1 Score treat feedback attributes generated on real submissions as the reference baseline.}
\label{tab:feedback_similarity}
\end{table}

\subsection*{Limitations of LLM-based Validation}
We explicitly highlight a key methodological limitation of our utility validation: the evaluation pipeline operates within an LLM-based computational ecosystem. Specifically, synthetic submissions were generated by LLMs, evaluated by ten commercial LLMs to produce feedback, and scored via an automated LLM-based feedback evaluator \citep{a:2}. 

This creates a potential ``hall of mirrors'' risk, wherein models may reward outputs that share common statistical regularities or stylistic biases inherent to LLM-generated text. While automated metrics provide a scalable and reproducible baseline, they cannot substitute for human pedagogical expertise. The reported alignment scores indicate computational consistency between real and synthetic data under automated assessment, but do not constitute verified pedagogical efficacy in authentic classroom settings.

\subsection*{Privacy Protection Verification}
We assessed the effectiveness of privacy filtering through ablation experiments using o4-mini to detect personal identifying information (Table~\ref{tab:privacy}). Under our evaluation protocol, privacy protection rates achieved:
\begin{enumerate}[label=\arabic*)]
  \item In-prompt protection only: 99.1\%
  \item Privacy gate filter only: 98.9\%
  \item Combined SAM framework (Prompt + Gate): 100\%
\end{enumerate}
In contrast, naïve mimicry without protection mechanisms achieved only 96.5\%.

\begin{table}[ht]
\centering
\resizebox{0.55\linewidth}{!}{%
\begin{tabular}{cccc}
\hline
\multirow{2}{*}{\textbf{Method}} &
  \multicolumn{2}{c}{\textbf{Protection Mechanism}} &
  \multirow{2}{*}{\textbf{\begin{tabular}[c]{@{}c@{}}Privacy Protection Rate\\ (Direct Identifiers)\end{tabular}}} \\ \cline{2-3}
                      & Prompt Protection & Gate Protection & \\ \hline
Naïve Mimicry         & \ding{55}          & \ding{55}        & 96.5\% \\ \hline
\multirow{3}{*}{SAM} & $\checkmark$       & $\checkmark$     & \textbf{100\%} \\
                      & $\checkmark$       & \ding{55}        & 99.1\% \\
                      & \ding{55}          & $\checkmark$     & 98.9\% \\ \hline
\end{tabular}%
}
\caption{Ablation study of privacy protection rates across different mechanism configurations.}
\label{tab:privacy}
\end{table}

\noindent\textbf{Scope and Boundary of Privacy Verification.} We emphasize that the ``100\%'' protection rate reported in Table~\ref{tab:privacy} strictly reflects the complete removal of \textit{explicit direct identifiers} (specifically student names, staff names, and numeric student IDs) as measured by the automated evaluation protocol across the tested sample. It does not constitute a formal mathematical guarantee (e.g., differential privacy) against all potential forms of indirect identifiers, unique personal anecdotes, idiosyncratic coding styles, or unmeasured quasi-identifiers. Furthermore, because the privacy gate relied on the same model family (o4-mini) as the generator, there remains a risk of shared model blind spots. We recommend that future deployments employ independent multi-model ensembles or human privacy audits for sensitive corpora.

\section*{Usage Notes}

The release of the SCALEFeedback dataset offers a standardized, open-access benchmark for exploring automated feedback generation, rubric adaptation, and feedback evaluation across 59 computer science subjects. However, researchers and practitioners should take note of several critical methodological boundaries and usage restrictions:

\noindent\textbf{1. Institutional and Disciplinary Context.} The dataset was derived from a single faculty of Information Technology at an Australian university (Monash University). It reflects the curriculum design, pedagogical norms, and English-language conventions of a Western higher education institution. Users should not assume that the dataset represents a global standard across different cultural, linguistic, or institutional contexts without domain-specific validation. Furthermore, the SAM framework has been validated solely on computer science coursework; extending this framework to open-ended, subjective disciplines (such as creative writing or the humanities) remains untested and speculative.

\noindent\textbf{2. Coursework Level and Sampling Bias.} Due to the 7,500-word context-length cutoff applied during data acquisition, the dataset primarily reflects foundational to intermediate computer science coursework (e.g., modular coding exercises, standard laboratory reports, database design tasks). It systematically lacks representation of large-scale, multi-file capstone projects, extensive software architecture portfolios, or multi-month team deliverables.

\noindent\textbf{3. Regression to the Mean and Restricted Use Cases.} As shown in Figure~\ref{fig:submission_violin} and our MAE breakdown, the synthetic generation process exhibits a smoothing effect, regressing toward typical and median performance tiers. The dataset lacks the full variance of extreme long-tail student behaviors, such as severely flawed conceptual submissions, bizarre formatting edge cases, or academic misconduct. \textbf{We strictly caution researchers against using this dataset for downstream tasks such as at-risk student detection, automated failing-grade classification, or academic integrity screening}, as these tasks depend on the exact tail-end anomalies that synthetic generation smooths away.

\noindent\textbf{4. Mitigation Strategies for Real-Classroom Deployment.} Practitioners building feedback models based on this dataset should not deploy them in live classrooms without proper edge-case mitigation. Feedback systems should incorporate domain-specific outlier detection, uncertainty estimation, and confidence-based routing mechanisms that automatically direct ambiguous, highly atypical, or severely failing submissions to human instructional staff.

\noindent\textbf{5. Absence of Symbolic and Execution-based Correctness Verification.} While synthetic submissions plausibly emulate student code structures and reasoning patterns for feedback prompting, the numerical and computational values in the synthetic tasks were generated probabilistically by LLMs without validation via symbolic solvers or sandbox code execution engines. The dataset should not be treated as a ground-truth execution benchmark for exact mathematical or runtime verification.

\noindent\textbf{6. Absence of Large-Scale Human Pedagogical Ground Truth.} Due to institutional privacy restrictions governing access to raw student submissions, utility validation relied on automated multi-model evaluations. While these demonstrate high computational consistency, future research should integrate expert educator evaluations to assess subtle pedagogical nuances, emotional tone, and actionable guidance.

\noindent\textbf{7. Risks of Model Collapse.} Downstream developers are advised to use SCALEFeedback primarily for benchmarking, prompt engineering, and diagnostic evaluation. We caution against utilizing this pre-smoothed synthetic corpus as the sole training or fine-tuning dataset for automated AI tutors, as training generative models recursively on smoothed synthetic text risks accelerating model collapse and degrading performance on diverse student populations.

\noindent\textbf{8. Privacy and Copyright Framing.} The SAM framework was developed with the objective of mitigating privacy risks and copyright infringement through synthetic imitation. While empirical testing demonstrated 100\% removal of explicit student identifiers, these measures reflect risk-mitigation engineering and should not be construed as absolute legal guarantees across differing multi-jurisdictional frameworks.

\section*{Data availability}

The complete SCALEFeedback synthetic dataset is openly available on the Open Science Framework (OSF) repository \citep{dataset} at \url{https://doi.org/10.17605/OSF.IO/9PWMX}. 

Due to ethical and legal privacy constraints governing student educational records, the original authentic assignment specifications and student submissions cannot be deposited in a public repository. To support reproducibility, access to de-identified raw reference data may be granted to qualified academic researchers for non-commercial verification purposes upon reasonable request. Inquiries should be directed to the Human Research Ethics project manager Guanliang Chen (\href{mailto:guanliang.chen@monash.edu}{guanliang.chen@monash.edu}), and lead author, Keyang Qian (\href{mailto:keyang.qian@monash.edu}{keyang.qian@monash.edu}). Access is subject to approval by the Monash University Human Research Ethics Committee and the execution of a formal Data Transfer Agreement (DTA).

\section*{Code availability}
The code used for synthetic data generation via the SAM framework, baseline comparisons, and statistical validation was implemented in Python 3.11. The complete codebase is publicly accessible on GitHub \citep{code} at \url{https://github.com/Yuugure-Yuu/Open-Synthetic-University-Assignment-Rubric-and-Submission-Dataset} under the open-source MIT license.

\bibliography{sample}

\section*{Acknowledgements}
This research was supported in part by the Australian Research Council (DP220101209, DP240100069) and the Jacobs Foundation (CELLA 2 CERES).

\section*{Author contributions statement}
K.Q. collected the authentic data, developed the SAM framework, conducted synthetic generation and validation experiments, and drafted the manuscript. D.G. and L.Y. conceptualized and supervised the research. K.Y., W.D., F.J., Y.C., R.G., S.N., Z.S., and G.C. contributed to data curation, methodology, and technical validation. All authors reviewed and approved the final manuscript.

\section*{Competing interests}
The authors declare no competing interests.

\end{document}